\documentclass{sig-alternate-05-2015}

\begin{document}

\setcopyright{acmcopyright}

\title{Real-time collision detection method for deformable bodies}

\numberofauthors{5} 
%
\author{
\alignauthor
Claudio Paglia\\
       \affaddr{Digital Science Center}
       \email{paglia23@digtsc.com}
\alignauthor
Guido Moretti\\
       \affaddr{Digital Science Center}\\
       \email{morettig@digtsc.com}
}

\maketitle
\begin{abstract}
This paper presents a real-time solution for collision detection between objects based on the physics properties. Traditional approaches on collision detection often rely on the geometric relationships that computing the intersections between polygons. Such technique is very computationally expensive when applied for deformable objects. As an alternative, we approximate the 3D mesh in an spherical surface implicitly. This allows us to perform a coarse-level collision detection at extremely fast speed. Then a dynamic programming based procedure is applied to identify the collision in fine details. Our method demonstrates better prevention to collision tunnelling and works more efficiently than the state-of-the-arts.
\end{abstract}

\printccsdesc



\section{Introduction}
Collision detection and resolution of importance, physics-based animation. Real-time response for many applications, depending on the potential collision geometry localization efficiency and computing the intersection of the polygon is essential to a large extent. Physics-based animation in digital entertainment, medicine and military simulation training and other fields have a wide range of applications \cite{15}\cite{09}. Interactive 3D graphics applications are a class of continuous queries user input and provide output by three-dimensional graphics application class. Technology used to implement these types of applications, so that other types of possible areas of application techniques such as virtual reality. Virtual reality applications, examples of simulation visualization environment interactive three-dimensional graphics applications.

However, sufficient attention, numerical robustness of the algorithm, resulting in the ball sometimes penetrate obstacles. In fact, there seems to be a number of powerful collision detection algorithm in the literature, despite their importance, computer graphics application. A comprehensive summary of these techniques can be found in ~\cite{01}. BVH space hash are the two most commonly used spatial data structure. Construction BVH, many types of boundaries may be used as raw OBB AABB \cite{02}\cite{03}, k-DOP \cite{04} balls. Based on the most BVH collision detection methods, the leaf node polygon intersection test is inevitable. Although the method is quick update BVH  \cite{05}\cite{06} proposed, BVH updated in each frame is still time consuming. For spatial hashing method, using uniform grid hashing collision deformable body and self-collision detection effectively. In \cite{07} proposed a hierarchical spatial hashing method. Pabst \cite{08}  provides a parallel solving spatial hashing. Our approach is to focus on the best level collision detection and resolution, which is compatible with hierarchical spatial data structures and their acceleration policies.

Grid-based model approach is another important abstract methods. In \cite{14}, the model is proposed based on surface extraction method. Then, Kwon, in \cite{10} to improve this method, it is more suitable. However, it is generated as a thin single layer mesh model also has limitations. In \cite{11} Quadros overcome this limitation. Mesh a CAD model is generated by any node is not inserted in the solid internal use Delaunay triangulation method. However, a disadvantage of this method is, not in the good way to ensure the quality of the surface. Surface extraction method based on face detection has also been widely used. In the \cite{12} first proposed a detection method (FAG) surface adjacent to the drawings based on the detection judge the distance between the face and attributes. These face-to, and then used to generate the surface. Lee \cite{13} Improved face detection method graph based to allow extension of the rules to join the edge surfaces. However, the limit is, when a topology model is complex, people face often fail to recognize.

In this paper, we propose a new technology that does not rely on polygon intersection tests. On the contrary, checks carried out include adaptive implicit catch surface structure, which is a faster alternative crash test. Existing tree balls or spheres package based structures, our method does not produce art work closely together in the initial stage, and to update each frame of the entire structure. Our dynamic basis for generating circumspheres potential impact on the local geometry of the triangle and material properties; without updating circumspheres per frame. Compared with other spherical representation, hidden circumspheres surface may represent the original surface tightly without using dense spherical structure. This is really useful to save on the cost of collision detection. The simulation framework of this deformable object ball collision method based on the widely used integration.

\section{Collision Detection Method }
Our approach is focused at a later stage, we divided it into two narrow sub-phases: Phase surrounded by balls and ball. For boundary ball stage, potential collision checked by a simple test ball border triangle primitives. When the possibility of collision is established, it moves to the ball stage, we tested basic dynamic elements generated by the circumscribed spheres overlap and using an external ball successfully resolve the conflict.

\subsection{Method Overview}
For the whole process of collision detection abstract method, as shown in Fig. Figure 1 is the original model, which is simple to use thin-walled ribs artificial model \cite{16}. The first step is to identify the rib features from the original model and the results shown in Figure 1. Then decomposed using ribs and tissues to produce a regional hierarchy. A film based on NURBS surfaces abstract as the main basic components, in the first stage, and shift operations for abstract ribfeatures
In other aspects. Finally, the surface models generated as shown in Figure 1.

\begin{figure}[htb]
\centerline{\epsfig{figure=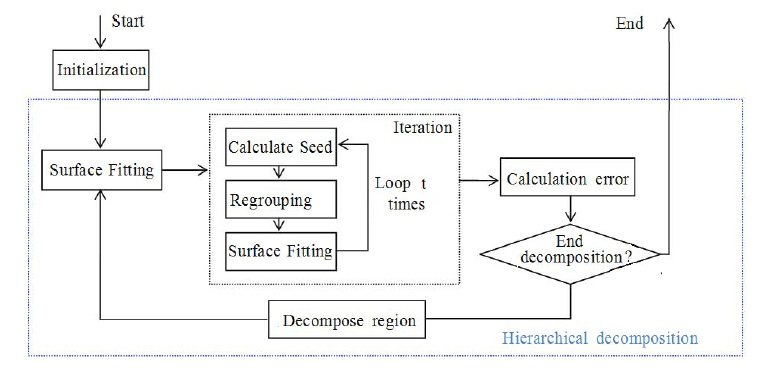, width=9cm}}
\caption{Overview of our method}
\end{figure}

In determining the potential impact of the bounding sphere stage, we calculated the catch perform intersection testing. According to the basic principles of geometry \cite{17}, the center of the ball can be calculated as:

\begin{equation}
x = x_d - Nd\phi\cdot R_c \cdot \cos(\theta)
\end{equation}

Where $R_d$ is the radius of the bounding sphere. In fact, not only for the triangular shape, and the local geometry of the surface influences the size of the external ball \cite{18}. Surface curvature is a good reflection of the local geometrical characteristics. Estimates of triangular mesh surface curvature can be used in three ways: repair method, method and curvature tensor averaging kinds of methods. In order to meet the needs of real-time response, we have adopted a dual grid computing efficiency  based on the approximation of the original structure of the curvature of the triangular mesh surface. We geometric center of the adjacent link up to build one pair of grid construction, and use of curvature
Dual mesh vertices approximate the original surface curvature corresponding. Gaussian curvature double hat mesh vertices used to approximate solutions.

\subsection{Spherical Surface Generation}
In order to facilitate the description of our approach, let us do some definitions. As seen in Figure 2, we define the polygonal surface covered ball coverage area. Area ball does not involve any potential external polygonal surface called non-covered surface. The original triangle normal line and the external sphere center and apex of the triangle is called to determine the angle between the security angle \cite{19}. From this perspective, the cone produced is called the normal safety cone.

\begin{figure}[htb]
\centerline{\epsfig{figure=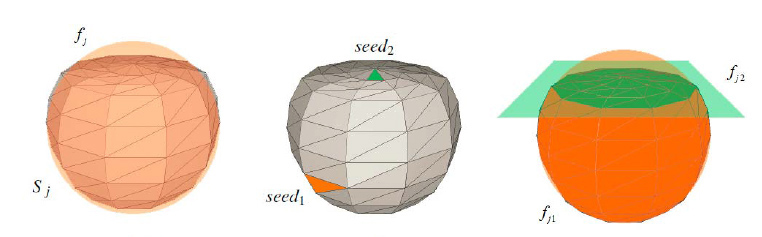, width=9cm}}
\caption{The method of polygon advantage in dealing with the problem of collisions based detection}
\end{figure}

However, the curvature may be zero, if the surface is a plane. Then the ball will be infinite because the formula R 4 is unlimited. In fact, if the surface is a plane, the curvature should not affect the size of the ball in. That is, in four equations, should be 1. From the perspective of collision detection, sharp area (large curvature) mesh more easily penetrate and tunnel artifacts due to their physical flexibility. This sharp areas, artifacts tolerance than good approximation, is more important. When K of the ball, the ball will be surrounded by just provide a good resistance artifacts. Meanwhile, such a ball can be adjusted to reduce unnecessary collision detection, as shown in Fig. Therefore, we use the curvature factor Hermite interpolation H3 (K) of. We assume that when R = 1 K > 0. We can set a threshold hthreshold K, when $k > hthreshold$, $H3 (K) = 0$. Cubic Hermite can be written as

\begin{equation}
C(p + \Delta p) = -\frac{C(P)}{\sum_\omega|\Delta_p C(p)^2|}
\end{equation}

In the XC is the corresponding triangle circumcenter, NC is the standardization of the surface and $\phi$ is a factor in the XC translation x- NC on a global scale. However, unnecessary external sphere intersection test will increase the size of the ball with an external evolution (see section 4.2). If we will be unified external dimension $\phi$  ball all external triangles approximate center of the ball does not mesh surface mesh surface and because of the distance between the center of the ball is uniform, as shown in Figure 2.

\begin{equation}
D= \frac{1}{3}\times(4 \times \pi + \sum_{i=0}^{k-1}\times \alpha_i)\times B
\end{equation}

I was my angle $\alpha$ between two consecutive edge connector $x$. It is a sum of n triangles share vertices my area $x$. Therefore, a close radius of the sphere in $x$: $R = \frac{1}{K}$ in order to affect the local geometry of the size of the ball, we should integrate the radius $R = \frac{1}{K}$  for the formula (2).

\subsection{Shape Update Process}
Dynamic simulation method to determine the impact of the collision response object. However, accurate collision response is calculated using a dynamic model of an expensive cost. In modern animation industry, visually reasonable and effective algorithm is the importance of real-time interactive applications. In this article, we use dynamic simulation framework based on location. In order to make the independence of this article, let us first review the dynamic workflow based on location.

And accelerating convergence stability analysis: Oscillation is a physics-based simulation common instability problem PBD. Currently PBD integrated physics engine (such as a bullet) will also be affected by this problem. Convergence is one of the reasons low speed oscillation. If there are various types of constraints to solve the current type is to limit possible violations previously addressed. Therefore, the iterative solver these constraints are necessary in order to find a stable converged solution. However, when using the original as a collision polygon, polygonal shape constraint changes brought about instability of the solution process. Unstable collision surface, due to the high variability of polygonal shape, particularly low quality grid, often caused by a collision constraints. For ball to ball collision, two circumspheres junctions may lead to erroneous collision detection, because the intersection
Probably in the area of non-covered surface. In order to distinguish the actual impact from the wrong collision, we use the method shown in Figure 7. Two collision surfaces (green and blue plane surface), if the red line (or blue face and determine the external surface of the green ball center) within a green ball of external safety cone, then we consider the collision, and vice versa.

\section{Results}

In the 2.1 update proposed threshold d, the accuracy of our testing for different d stability and efficiency (see Figure 2). When d = 0, the ball will lead to instability and inefficiency updated every frame. The cause of instability is the changing size of the increasing instability of the ball about the solving process. Inefficient because of the ball in the calculation of each frame. Relative error is because with $\delta$ approximation becomes inaccurate. Great ¦Ä, the external sphere updates require less time. Extreme situation is not updated external ball. Then deforming meshes external sphere approximation is the worst, causing instability. When $\delta$ is less than about $70$ percent, the stability improved by $\delta$.

\begin{figure}[htb]
\centerline{\epsfig{figure=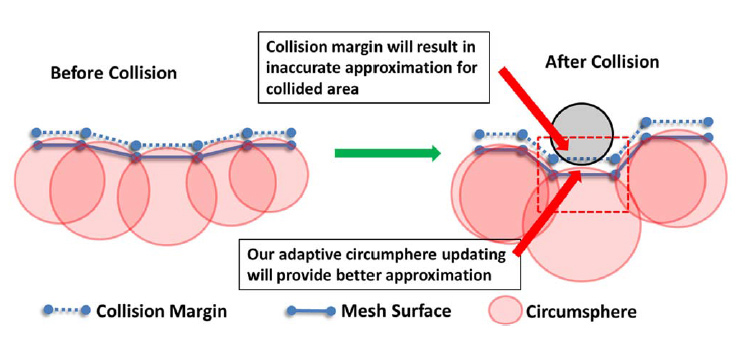, width=9cm}}
\caption{Comparison of visual effect is evident polygon-based method and boundary sphere method}
\end{figure}

For the purpose of simulation of different types, different update rates should be used (as shown in Figure 16). Precision-oriented site, it is a virtual surgery system, which contains detailed collision and sliding. d should be as small as possible, within a stable range ($30\% ~ 200\%$). For the stable scene, it contains colliding primitives between the thin cloth and various. To maintain stability and accuracy should be used $d = 70\%$. For efficiency scenario, which involves sliding and impact of various modifications detailed objects and different collision primitives between. Maintain efficiency and stability, we have $d = 90\%$

\begin{figure}[htb]
\centerline{\epsfig{figure=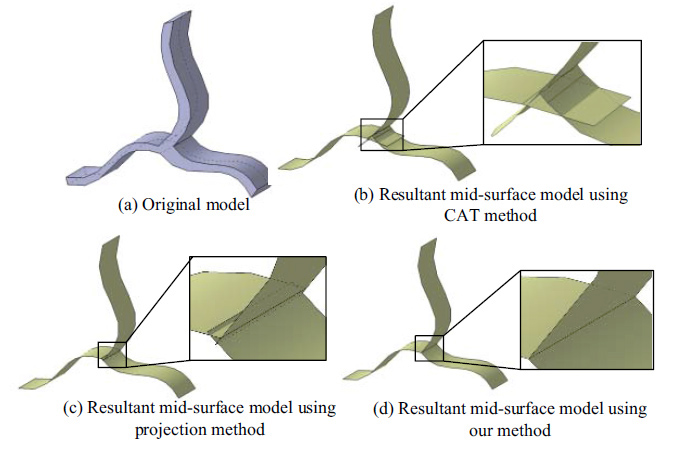, width=9cm}}
\caption{For testing accuracy, stability and efficiency of the scene}
\end{figure}

Collision detection in each frame and decomposition of the time consumed is a good measure of the efficiency of the algorithm. Boundary ball method is the most effective way, because it only needs to be updated in the center, radius and overlap test execution, collision. We need a method to recalculate change the shape of the original catch potential exceeds a given threshold. Polygon Intersection method is the most time-consuming, because in every frame, every frame once a polygon intersection. Fluctuation vertex position is a reflection of stability. Thus, the average position of all collisions we measured variation vertices in each frame, in order to assess the stability. From Fig. 13, polygons, and methods bounding sphere is subject to volatility, the number of iterations is small. For the method of polygons, polygon shapes high variability is the main reason for instability. The method for bounding sphere of instability is the difference between its approximate mesh surface caused. Our sphere of external structure, on the contrary, not only to provide a good approximation of the mesh surface, but also increases the instability less constraint solving process, so that our approach is particularly low performance best iterations.

\begin{figure}[htb]
\centerline{\epsfig{figure=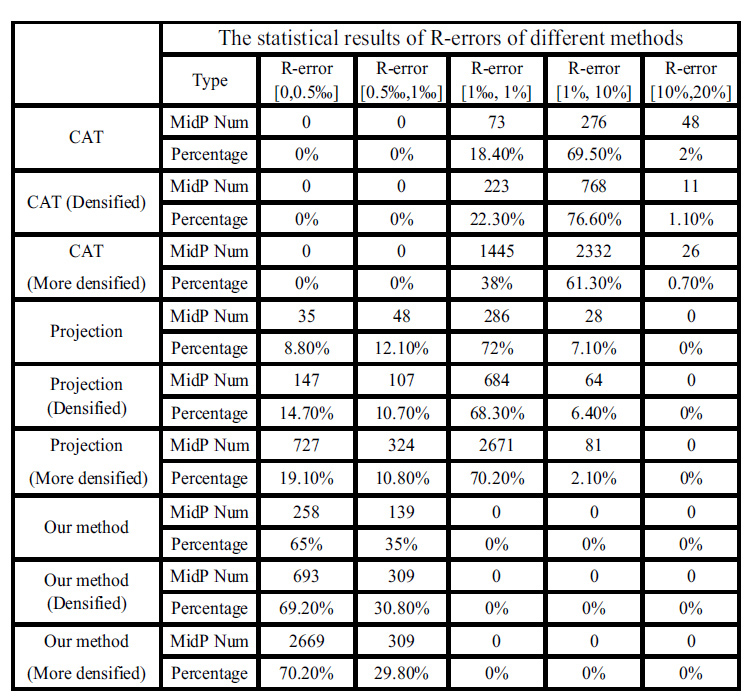, width=9cm}}
\caption{Accuracy comparison under different methods}
\end{figure}

\section{Conclusions}
Our algorithm has been implemented and tested a variety of models and design parameters. Our results show that our method can provide a fairly satisfactory rendering speed and a physical deformation and rigid body vertexes reasonable progress and responses between the cadmium up to 3000, and the original number up to 100,000. Future work will focus on the study of the deformation and cadmium fluid and deformation models between cadmium and response. We have introduced a dynamic deformation of objects collision and self-collision detection method. Any objects instead of calculating global bounding box, and explicitly the subdivision of space, we recommend using a hash function to map a three-dimensional cell hash table, enabling a very efficient, fine implicit space Minute. In the tetrahedron test is based on the actual vertex coordinates of the center of gravity. It provides information that can be used for collision physics is based.

\bibliographystyle{ieeetr}
\bibliography{paper}

\end{document}